\documentclass[twocolumn,showpacs,preprintnumbers,amsmath,amssymb]{revtex4}

\usepackage{graphicx}
\usepackage{dcolumn}
\usepackage{bm}

\begin{document}

\preprint{R. Palai {\it et al}.}

\title{Phonon Spectroscopy Near Phase Transition Temperatures in
Multferroic BiFeO$_{3}$ Epitaxial Thin Films} 

\author{R. Palai$^{1}$, J.F. Scott$^{2}$, and R.S. Katiyar$^{1}$}
\affiliation{$^{1}$Department of Physics and Institute for Functional Nanomaterials,
University of Puerto Rico, San Juan, PR 00931-3343, USA\\
$^{2}$Department of Physics, Cavendish Laboratory, University of
Cambridge, Cambridge, CB3 0HE, UK}


\date{\today}

\renewcommand{\baselinestretch}{} 
\newcommand{\bfo}{BiFeO$_{3}$}
\newcommand{\sto}{SrTiO$_{3}$}
\newcommand{\sro}{SrRuO$_{3}$}
\newcommand{\cm}{cm$^{\rm -1}$}
\newcommand{\tn}{$T_{\rm N}$}
\newcommand{\pr}{$P_{\rm r}$}
\newcommand{\mr}{$M_{\rm r}$}
\newcommand{\tc}{$T_{\rm c}$}
\newcommand{\jc}{$J_{\rm c}$}
\newcommand{\rs}{$R^{\rm 2}$}
\newcommand{\dc}{$^{\circ}$C}
\newcommand{\rr}{$r_{\rm R}$}

\begin{abstract}
We report a Raman scattering investigation of multiferroic bismuth
ferrite (\bfo) epitaxial ($c$-axis oriented) thin films from -192
to 1000\dc. Phonon anomalies have been observed in three
temperature regions: in the $\gamma$-phase from 930\dc\ to 950\dc;
at $\sim$ 370\dc, N$\acute{e}$el temperature (\tn), and at
$\sim$~-123\dc, due to a phase transition of unknown type
(magnetic or structural). An attempt has been made to understand
the origin of the weak phonon-magnon coupling and the dynamics of
the phase sequence. The disappearance of several Raman modes at
$\sim$ 820\dc\ (\tc) is compatible with the known structural phase
transition and the Pbnm orthoferrite space group assigned by
Arnold {\it et al.} \cite{arnold:09}. The spectra also revealed a
{\it non-cubic} $\beta$-phase from 820-930\dc\ and the same {\it
non-cubic} phase extends through the $\gamma$-phase between
930-950\dc, in agreement with Arnold {\it et al.}
\cite{arnold2:09}, and an evidence of a cubic $\delta$-phase
around 1000\dc\ in thin films that is not stable in powder and
bulk. Such a cubic phase has been theoretically predicted in
\cite{vasquez:prb09}. Micro-Raman scattering and X-ray diffraction
showed no structural decomposition in thin films during the
thermal cycling from 22-1000\dc.

\end{abstract}

\pacs{77.55.+f, 78.30.-j, 77.80.Bh, 78.66.Nk}
\maketitle
\section{INTRODUCTION}

Multiferroics are the materials which display a coexistence of at
least two of the switchable states:  polarization, magnetization
or strain in the same phase \cite{hans:fe92}. In addition, they
may also exhibit a magnetoelectric (ME) effect: magnetization
induced by an electric field and electric polarization by means of
magnetic field \cite{fiebig:nature1}. The current interest in
multiferroics is largely based on engineered epitaxial and
heterostructured thin films, because their physical properties are
as good as bulk and permit technological applications in data
storage, magnetic recording, spintronics, quantum electromagnets,
and sensors \cite{wang:science1,tokura:science06, scott:nm07}.
Devices made up of multiferroic materials can perform more than
one task and facilitate device miniaturization. A weak ME effect
has been observed in most multiferroics, generally showing a small
change in their spontaneous polarization under applied magnetic
field \cite{fiebig:nature1,zhao:nm06,chu:am076}. However, the
complete switching of ferroelectric domains by applied magnetic
fields has rarely been observed. Why and under what circumstances
a large coupling should exist and how to control the coupling are
still open questions. Understanding the physics of the different
possible interactions between magnetic and electric order
parameters {\it i.e.} giving rise to magnetoelectric (ME) coupling
would be very useful.

Magnetism and ferroelectricity are involved with local spin
ordering and off-center structural distortions, respectively
\cite{spaldin:jpcb}. These are quite complementary phenomena that
coexist in certain multiferroic materials. Currently, \bfo (BFO)
is one of the most widely studied multiferroics because it is one
of only two or three single-phase multiferroics at room
temperature $i.e$ an antiferromagnetic (AFM) incommensurate phase
with cycloidal modulation ($\lambda$ $\approx$ 60 nm) below
$\approx$~370 \dc\ \cite{sosnowska:jpc82, smolenkii:SPU82},
ferroelectric up to $\approx$~820\dc\ \cite{hans:fe84}, and
ferroelastic between 820-930\dc\ \cite{palai:07}. Bulk BFO
crystallizes in a rhombohedral ($a$~=~5.58~\AA\ and
$\alpha$~=~89.5$^{0}$) structure at room temperature (RT) with
space group R3c (C$^{6}_{3v}$) and antiferromagnetism of $G$-type
\cite{smolenkii:SPU82,hans:ac90}. The structure and properties of
bulk BFO have been studied extensively \cite{smolenkii:SPU82,
hans:ac90, bucci:jpc72,polomska:physlett74} and although early
values of polarization were low (\pr~=~6.1~$\mu$C/cm$^{2}$) due to
sample quality, \pr~=~40-100~$\mu$C/cm$^{2}$ was recently found in
bulk by several different groups \cite{shuartsman:apl07,
lebeugle:prb07}. The epitaxially grown thin films of BFO on
SrTiO$_{3}$ (STO) substrates show very high values of \pr ($\sim$
100 $\mu$C/cm$^{2}$) \cite{wang:science1} compared to the best
known ferroelectrics such as PbZrTiO$_{3}$ ($\sim$
70$\mu$C/cm$^{2}$) and BaSrTiO$_{3}$ ($\sim$ 30$\mu$C/cm$^{2}$).
This makes BFO a potential material for novel device applications.

The motivation for the present study is manifold. The first
objective is to test the recent space group determination of the
$\gamma$-phase reported by Arnold {\it et al.} \cite{arnold:09} as
being orthorhombic.  Their definitive neutron study showed that
the $\gamma$-phase is indeed stable (which in itself had been
controversial), and that it has the same orthorhombic Pbnm
orthoferrite symmetry as does the $\beta$-phase.  A cubic Pm3m
perovskite structure was definitely ruled out, although a body-
centered orthorhombic space group was indistinguishable from the
primitive Pbnm.  A main aim of our Raman study is to test the
orthorhombic crystal class for the $\gamma$-phase and see whether
we can further distinguish between primitive and body-centered
orthorhombic structures.

\section{EXPERIMENTAL DETAILS}
We investigated 300~nm (001) BFO thin films on STO (100)
substrates with $\sim$ 25 nm thick SrRuO$_{3}$ (SRO) buffer layer
by pulsed laser deposition (PLD). A Jovin Yvon T64000 micro-Raman
microprobe system with Ar ion laser ($\lambda$ = 514.5 nm) in
backscattering geometry was used for polarized and temperature
dependent Raman scattering. Sample deposition and experimental
details are given in \cite{palai:07}.

The X-ray diffraction (XRD) pattern (see Fig.~\ref{xrd}a) of the
BFO films taken using CuK$_\alpha$ (1.5406\AA) radiation  show
$c$-axis (pseudo-cubic $<$001$>$ direction perpendicular to the
substrate) orientation with a high degree of crystallinity. The
$c$-axis length was found to be 3.95~\AA, which implies epitaxial
strain is quite relaxed. This agrees with the reported values
($c$~=~3.997~\AA) \cite{xu:apl05}.

The comparison of the unpolarized (perpendicular to the $<$001$>$
of the substrate) Raman spectrum of BFO thin film with STO and
SRO/STO spectra (cf. Fig. 2a in \cite{palai:07}) precludes any
Raman contribution from the substrate and bottom electrode; to the
contrary, we observed a dip, rather than a peak, at the STO
strongest peak position.  As is evident from the intensity
comparison, all of these peaks are due to the BFO normal modes of
vibrations and none of them arose from the substrate. We verified
our results using target materials, single crystals, and also by
growing (001) BFO films on different substrates.
\begin{figure*}
\begin{center}
\includegraphics [width=0.4\textwidth,clip]{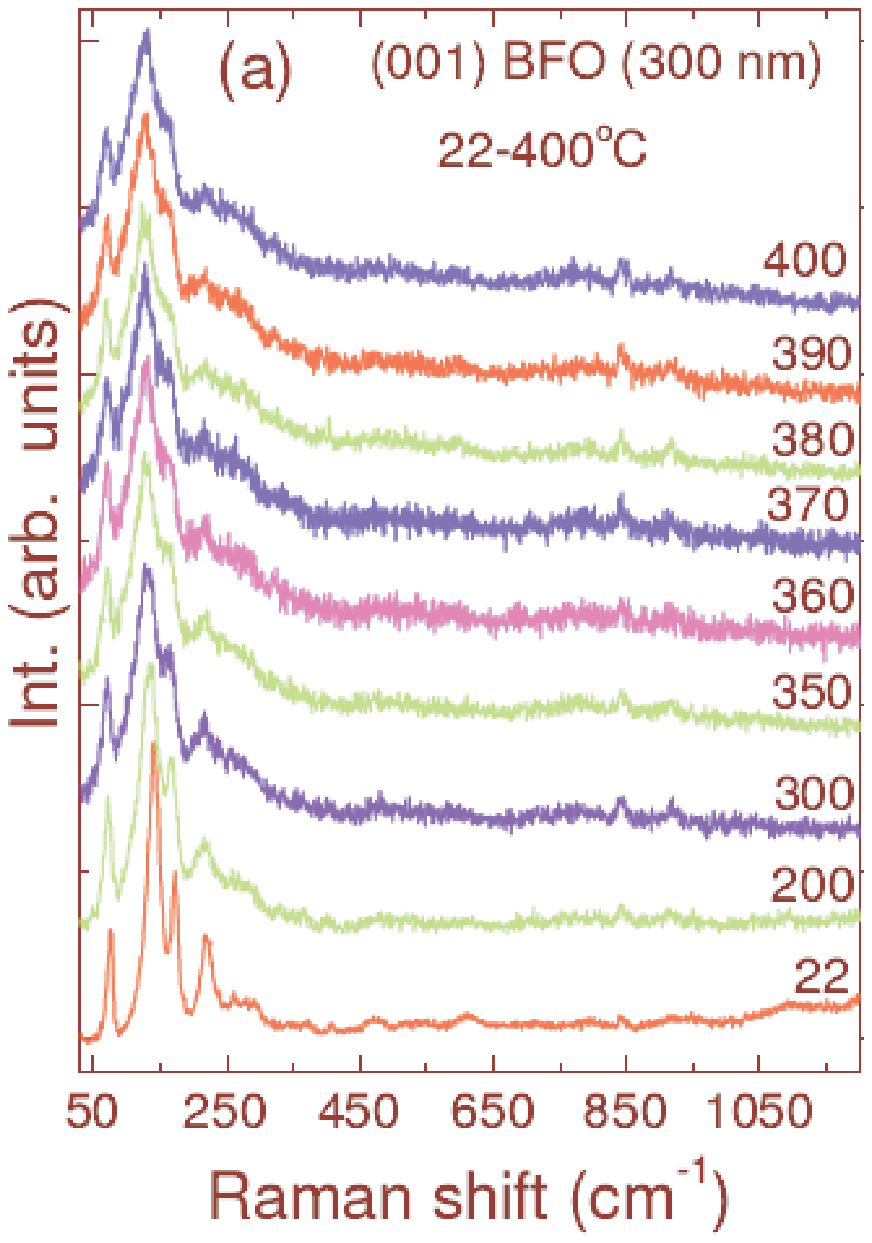}
\includegraphics
[width=0.4\textwidth,clip]{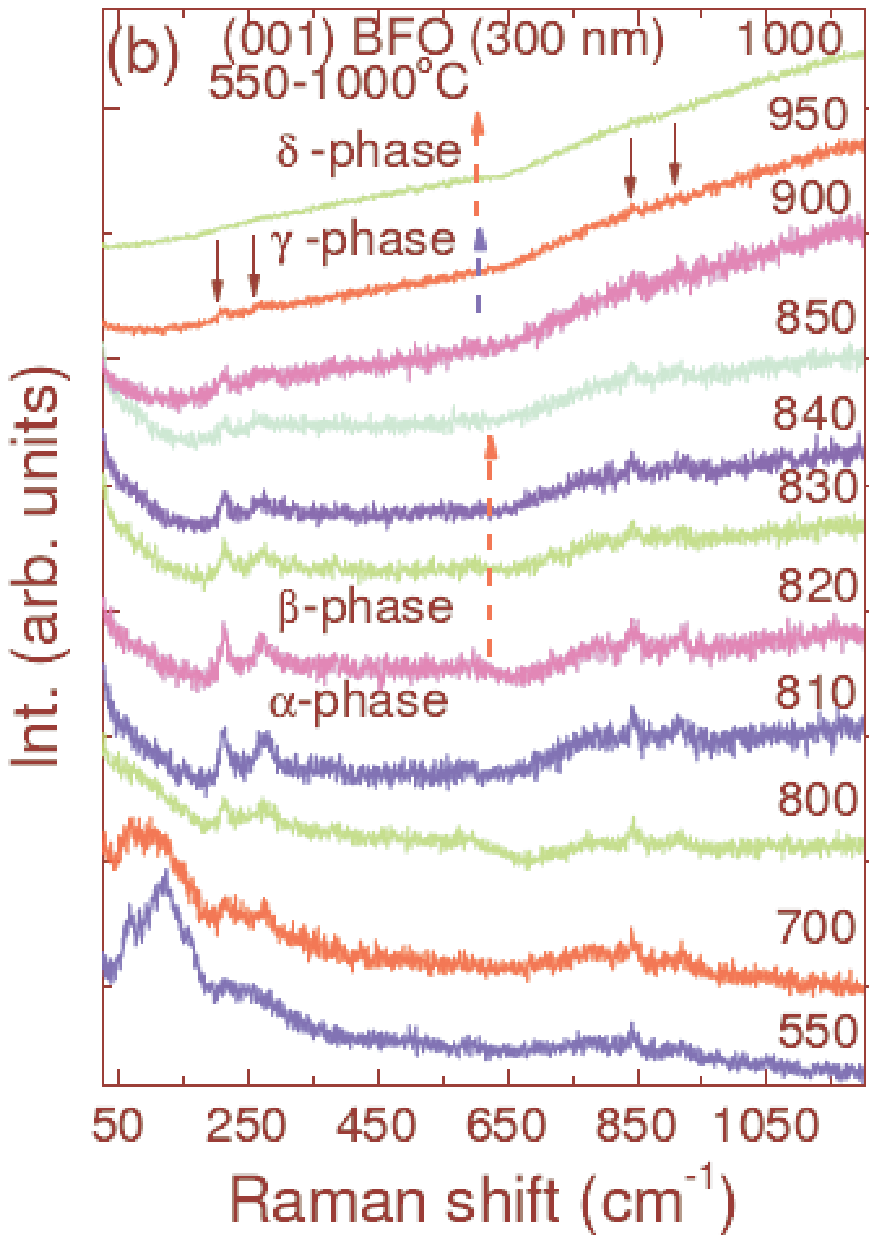}\vspace{0.1in}
\includegraphics [width=0.43\textwidth,clip]{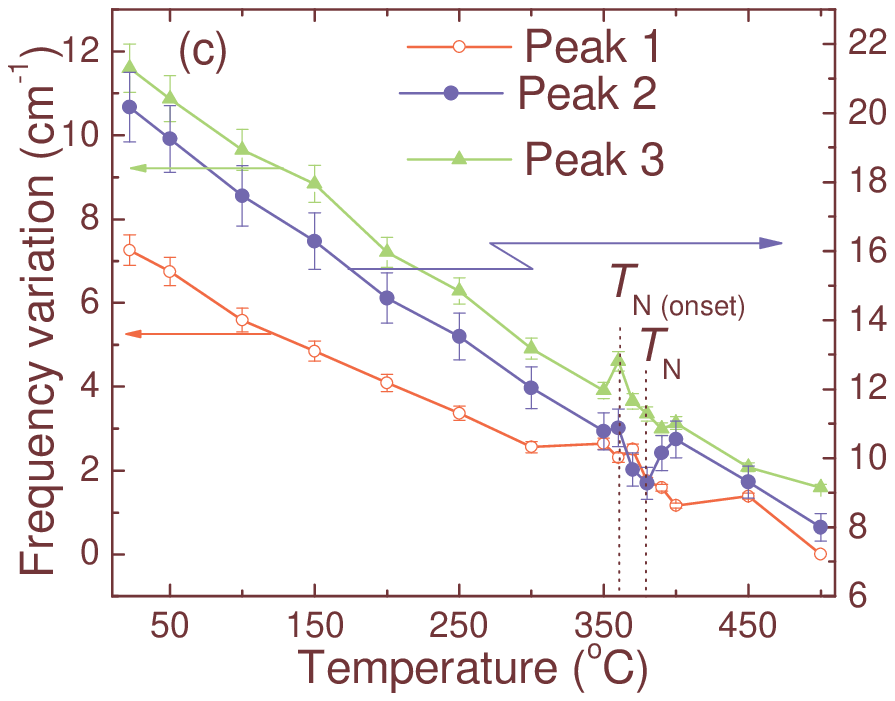}
\includegraphics [width=0.4\textwidth,clip]{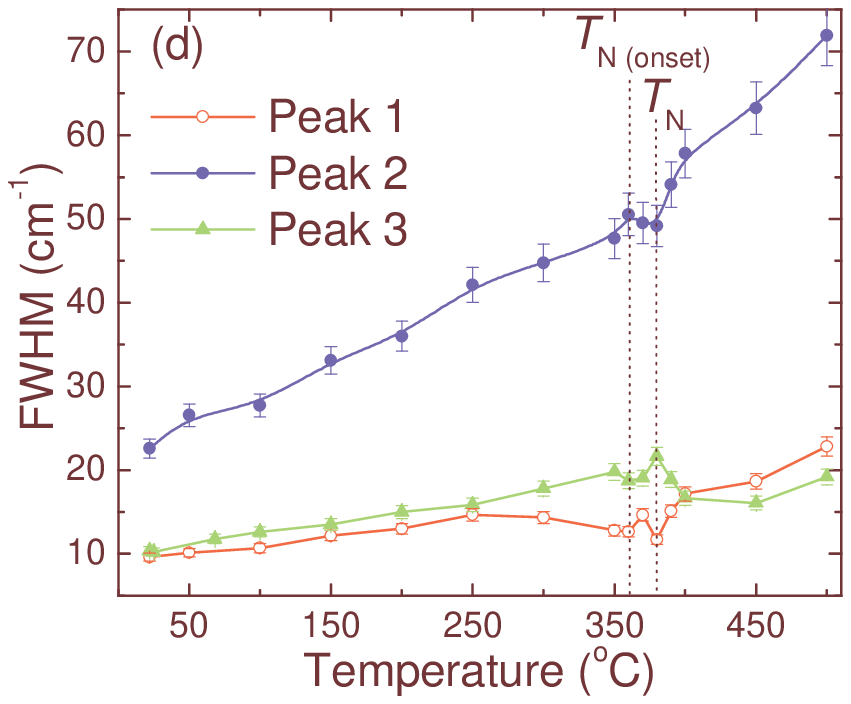}
\caption{\sf  Temperature dependent Raman spectra of (001)BFO film
on SRO/STO from 22-400\dc\ (a) and 550-1000\dc\ (b). The graphs in
(b) were adapted from our earlier work \cite{palai:07}. The
beginning of the {\it dashed arrows} pointing up in (b) shows the
beginning of the new phase. The existence of phonons in
$\gamma$-phase in (b) are marked with the solid arrows pointing
down. The $\alpha$-phase extends up to 820\dc; (c) and (d)
temperature dependence (RT-500\dc) of phonon frequencies variation
and FWHM for 72 (peak 1), 140 (peak 2), and 171(peak 3) \cm,
respectively.} \label{raman:ht}
\end{center}
\end{figure*}
\section{RESULTS AND DISCUSSION}
\subsection{Phonons in the $\gamma$-phase}

Fig~.\ref{raman:ht} shows the temperature dependent Raman spectra
of (001)BFO film (300nm thick) on \sto\ substrates with \sro\
buffer layer (25nm thick). As can be seen from
Fig~.\ref{raman:ht}b, four phonon features marked with arrows
persist into the $\gamma$-phase from 930-950\dc. Therefore this
confirms an existence of non-cubic $\gamma$-phase. No first-order
Raman scattering from phonons is allowed for cubic Pm3m (because
each ion is at an inversion center, all phonons are odd-parity).
This is in agreement with the observation of an orthorhombic
symmetry for the $\gamma$-phase by Arnold {\it et al.}
\cite{arnold:09}. Although the Raman lines are broad and weak at
these temperatures, they exhibit no significant frequency shift at
the $\beta$-$\gamma$ transition at 930\dc; nor are there
additional lines above or below 930\dc. Therefore it is very
likely that the structure is Pbnm in both $\beta$ and $\gamma$
phases and that no primitive-to-body-centered phase change occurs.
Note that Fig~.\ref{raman:ht} is for a thin film. Therefore we
conclude that the films do not differ from single crystals and
powders previously studied with regard to the $\gamma$-phase.

There is renewed interest in this $\beta$-$\gamma$ transition
because of the earlier discovery of a metal-insulator transition
at 931\dc\ where the orthorhombic-cubic transition takes place in
bulk \cite{palai:07} and that the same transition may occur at 47
GPa at RT. Earlier M$\ddot{\rm o}$ssbauer studies established that
at this pressure the magnetization also disappears
\cite{gavriliuk:JETP05}. Although Gavriliuk {\it et al.}
\cite{gavriliuk:JETP05} concluded that this is a
rhombohedral-rhombohedral symmetry-preserving Mott transition,
that seems quite unlikely, because Haumont {\it et al.}
\cite{haumont:prb09} have found several phase transitions at lower
pressure. Thus the symmetry of \bfo\ is not rhombohedral on either
side of the high-pressure metal-insulator transition. Whether the
transition is a Mott transition or a band transition is unproven.
Various theoretical models disagree: Vasquez {\it et al.}
\cite{vasquez:prb09}  get an {\it ab initio} Mott transition;
Clark {\it et al.} \cite{clark:apl07} got a band transition to a
semimetal from a screened exchange model.

Figs.~\ref{raman:ht}a and b show the temperature variation (from
RT up to 1000\dc) of unpolarized Raman spectra of a BFO (001) thin
film. A closer observation near the phase transitions reveals two
noticeable changes in the signature of the Raman spectra: the
disappearance of several stronger modes at $\sim$ 820~\dc\ and the
complete disappearance of all the modes  above 1000~\dc. This
temperature behavior implies that BFO maintains its room
temperature structure up to $\sim$ 820~\dc, indicating the a
structural (ferroelectric) phase transition, in agreement with the
earlier investigations on BFO bulk single crystal and
polycrystalline samples \cite{haumont:prb06}. Note that thin films
of BFO show first order phase transitions as in bulk, whereas STO
and PbTiO$_{3}$ PTO are known to be first order in bulk but second
order in thin films \cite{catalan:prl06,he:prl05}.

The presence of the four peaks ($\sim$ 213, 272, 820 and 918 \cm)
above 820\dc\ up to $\sim$ 950\dc\ (Fig.~\ref{raman:ht}b) shows
that the intermediate  beta-phase is {\it not} cubic
(P$m\bar{3}m$) as reported by Haumont {\it et al}
\cite{haumont:prb06}. In fact, the phase diagram of BFO
\cite{Speranskaya:izv65}, and its more recent revised versions
\cite{palai:07}, show that BFO possesses a non-cubic $\beta$-phase
between 820 to 933\dc\ before it goes to the $\gamma$-phase, and
the $\beta$-phase was recently shown to be orthorhombic by using
high temperature X-ray diffraction and domain structures
\cite{palai:07} and neutron diffraction \cite{arnold:09}. The
complete disappearance of peaks at above 950\dc\ -- {\it not at
930\dc} -- indicates that the $\gamma$ high temperature phase also
cannot be cubic (P$m\bar{3}m$), for which any first-order Raman
scattering is forbidden.

\begin{figure}
\begin{center}
\includegraphics [width=0.45\textwidth,clip]{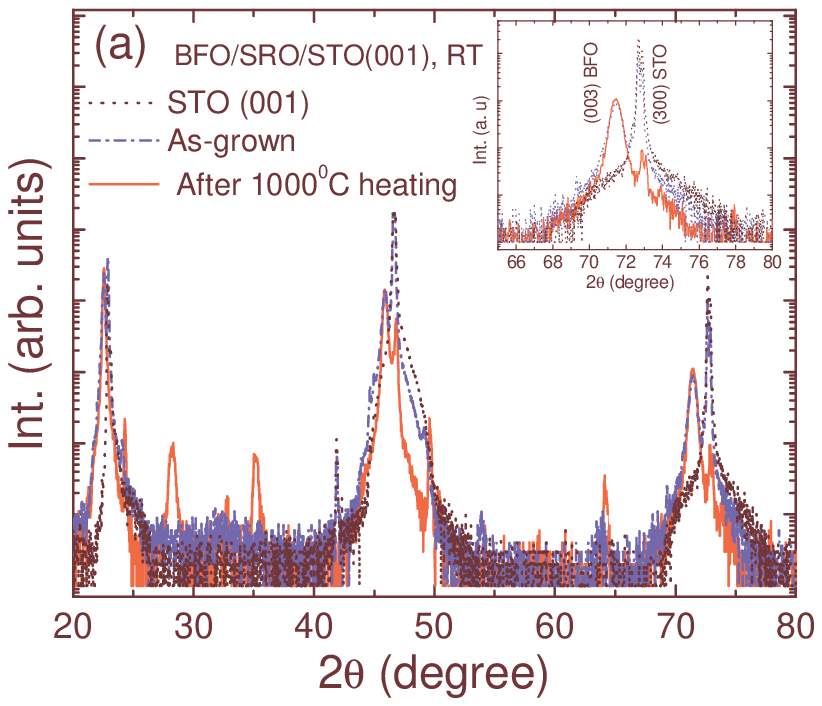}
\includegraphics [width=0.45\textwidth,clip]{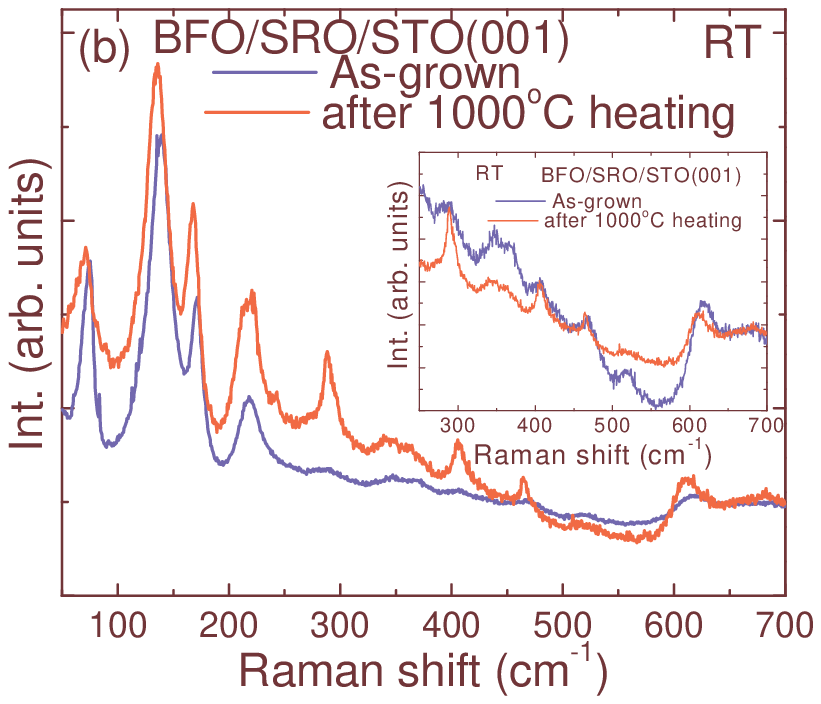}
\caption{\sf Room-temperature XRD patterns taken using
CuK$_\alpha$ (1.5406\AA) radiation (a) and  Raman spectra (b) of
(001)BFO film on SRO/STO before and after thermal cycle up to
1000\dc. XRD pattern of STO is given for comparison.} \label{xrd}
\end{center}
\end{figure}

Fig.~\ref{xrd}a shows the room-temperature XRD  patterns of an
as-grown film and film after it underwent 1000\dc\ thermal cycle.
As can be seen, the as-grown film is highly epitaxial showing only
(00l) peaks and became polycrystalline after thermal cycling. In
principle it is possible that the specimen would melt at high
temperatures and then recrystallize in the specimen holder (bottom
of the Pt crucible). However, we monitored the sample surface
continuously with an optical microscope and no thermal
decomposition was observed up to 1000\dc.

The Raman spectra (Fig.~\ref{xrd}b) before and after heating show
{\it exactly} same number of phonon modes, indicating either no
decomposition up to 1000\dc\ or complete recrystallization,
contrary to earlier studies \cite{bucci:jpc72}, which could be due
to the reduced surface/volume ratio, minimal surface imperfections
and defects, and increased stabilization from the substrate. Note
that the possibilities of subtle structural changes (small changes
on angles and/or in-plane lattice parameters) cannot be completely
ruled out. However, the Raman frequencies before and after thermal
cycling remain unchanged makes this unlikely. This fact favors
films over bulk or powder samples for very high temperature
studies in the future. Reaching the tetragonal and cubic phases
extrapolated from the powder study of Arnold et al.
\cite{arnold2:09} does not seem impossible with thin films.

\subsection{Phonon anomalies near \tn:}

There are discrepancies in the literature regarding both the
crystal structure of (001) BFO thin films {\it e.g.} with several
reports claiming tetragonal \cite{wang:science1,singh:prb06},
rhombohedral \cite{das:apl06, qi:apl05}, and monoclinic
\cite{xu:apl05, li:apl04} structure, and its phonons. Of
particular interest regarding phonon-magnon coupling in BFO was
the report \cite{haumont:prb06} of a very large (40 \cm) change in
the frequency of one long wavelength phonon branch near \tn.  We
emphasize in the present work that we see no such phenomenon
(Fig.~\ref{raman:ht}a). Instead we see in Figs.\ref{raman:ht}c and
d very small changes in frequency (1 or 2 \cm) and linewidth of
several polar modes, and we model them according to the
non-mean-field theory of Nugroho {\it et al} \cite{nugroho:prb07}.

In general, there are three kinds of phonon anomalies observed
near the phase transition temperatures:  a sigmoidal S-shaped
change in frequency (such as that reported by Haumont {\it et al}.
\cite{haumont:prb06} but not found in our work); a step
discontinuity; or a small ``bump" (increase) that returns to the
background level a few degrees above or below the transition
temperature.

In order to study the evolution of Raman signature around the
antiferromagnetic-paramagnetic (AFM-PM) phase transition, we
followed very closely the temperature dependence of few intense
phonon modes {\it i.e} 72 (Peak 1, we considered it as single peak
for the simplicity), 140 (Peak 2), and 171 (Peak 3) \cm\ from RT
up to 500\dc (Fig.~\ref{raman:ht}a). In general, the change in
phonon frequency band and width with temperature can be caused by
several factors, such as anharmonic scattering, spin-phonon
coupling, lattice expansion and/or contraction due to
anharmonicity and/or magnetostriction effects, and phonon
renormalization resulting from electron-phonon coupling
\cite{grando:prb99}. The latter one is not applicable here as BFO
is a highly resistive material, and the carrier concentration is
low. The change in ionic binding energies with temperature also
affects the change in phonon bandwidth in ionic compounds.
However, this is not applicable here as BFO insulating.
Figs.~\ref{raman:ht}c and d reveal the fluctuation of phonon
frequency and full-width at half maxima (FWHM) around 370\dc\
(onset could be $\sim$360\dc), which happens to be the \tn\ of
BFO. Near \tn we see small (1 or 2 \cm) changes in both peak
frequency and linewidth for three phonon modes at 72, 140, and 171
\cm. These satisfy the non-mean-field predictions of Nugroho {\it
et al} \cite{nugroho:prb07}. This behavior could be the
manifestation of phonon-magnon interaction that vanishes above
\tn, 370\dc. This weak interaction can be explained with the fact
that the magnetic phase transition is not accompanied by a
structural phase change. The observation of a rather weak
phonon-magnon interaction is also consistent with the theoretical
prediction of weak magnetization and ME coupling in BFO thin film
by Ederer {\it et al} \cite{ederer:prb05a}.

\subsection{Phonon anomalies near 140-150K:}

\begin{figure*}
\begin{center}
\includegraphics [width=0.35\textwidth,clip]{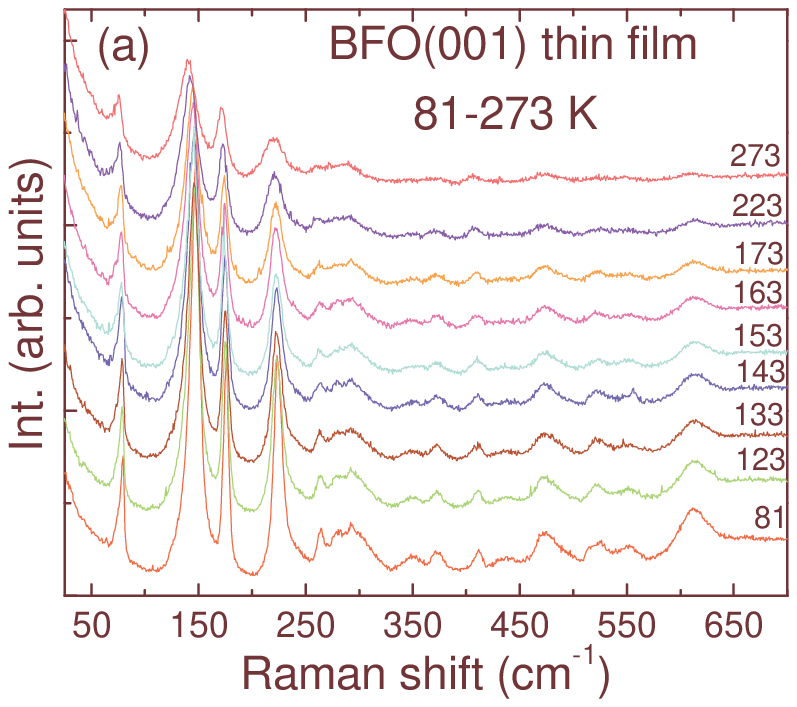}
\includegraphics [width=0.3\textwidth,clip]{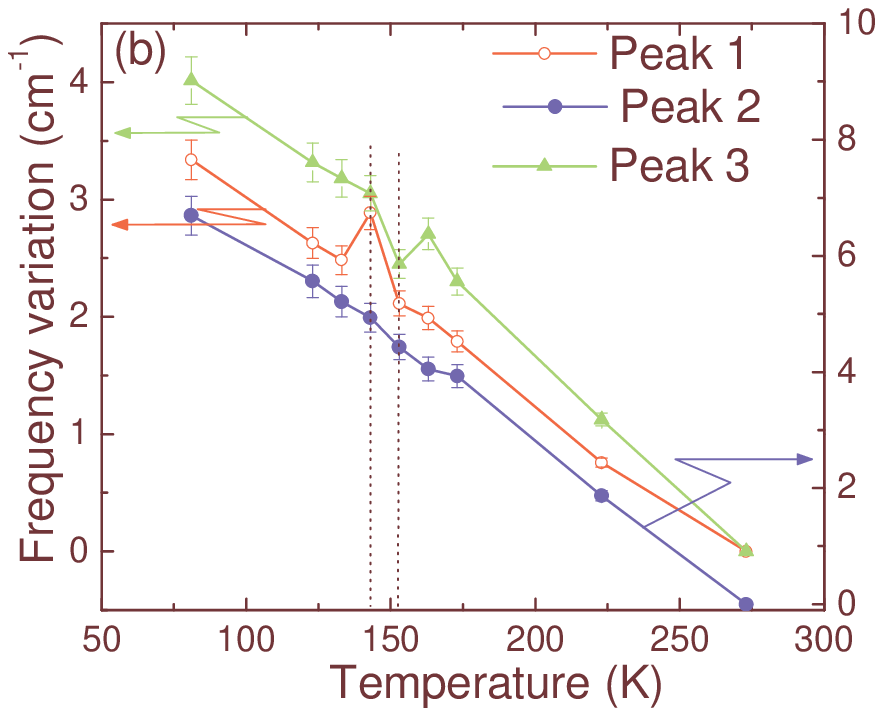}
\includegraphics [width=0.3\textwidth,clip]{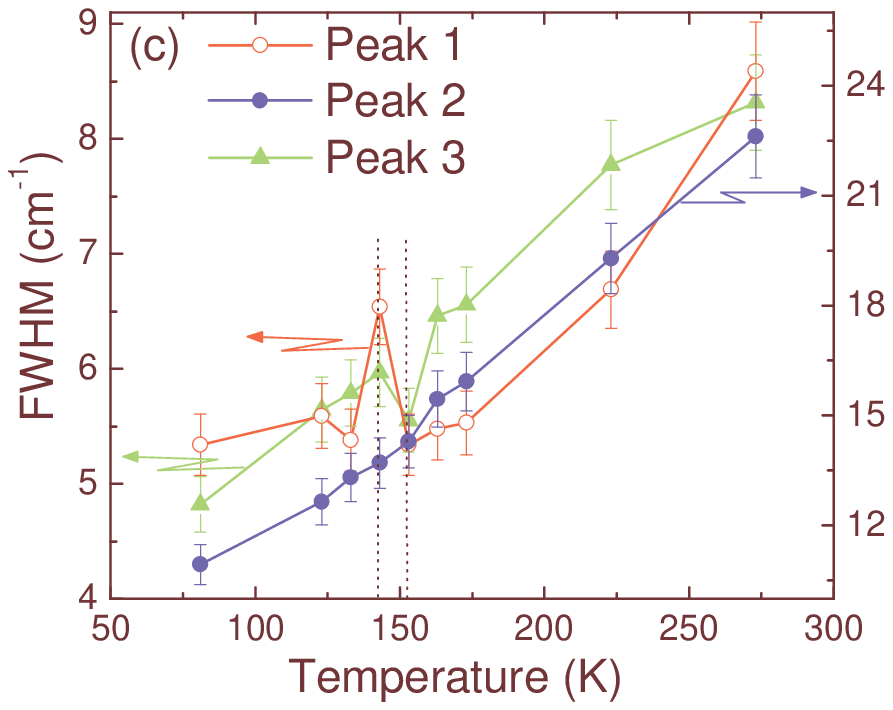}
\caption{\sf (a) Temperature dependent Raman spectra of (001)BFO
film on SRO/STO from 81K-RT; (b) and (c) temperature dependence of
phonon frequencies variation and FWHM for 72 (peak 1), 140 (peak
2), and 171 (peak 3)\cm\ down to 81K from RT, respectively. }
\label{raman:lt}
\end{center}
\end{figure*}

In order to study the cryogenic behavior of BFO thin films, we
carried out scattering measurements (Fig.~\ref{raman:lt}a) down to
81~K; no significant change in Raman spectra has been observed,
indicating the RT structure remained unchanged down to ~81K.
However, a close observation shows phonon anomalies around ~150K
(Figs.~\ref{raman:lt}b and c). This agrees with the observation of
change in magnetic order at ~150K by Pradhan {\it et al}.
\cite{pradhan:apl05}, but the nature of this phase transition
remains moot.

Figs.~\ref{raman:lt}b and c show small anomalies in the frequency
and linewidth of two phonon branches at 140 and 171 \cm near 140K,
a temperature at which anomalies have previously been reported.
Although the changes are small and the data sparse, they are
highly reproducible.  We are aware of similar observations on
single crystals by Brahim Dkhil \cite{dkhil:p09}, and we thank him
for preprints of his work.  The present data merely show that the
same effects are present in thin films.

The nature of the phase transition at 140K remains unknown.  There
are anomalies in magnon scattering cross-sections
\cite{catalan:am09,singh:jpcm09} and linewidth \cite{scott:jpcm08,
scott:jpcm08b}, in mechanical loss tangent \cite{catalan:am09,
redfern:jpcm08} and 140K is the end-point in Almeida-Thouless data
plots \cite{scott:jpcm08,scott:jpcm08b}. However, the earlier
suggestion by our group \cite{singh:jpcm08} that 140K is a
spin-reorientation transition temperature is not confirmed by very
recent neutron scattering studies \cite{catalan:up09}  and
spin-glass effects\cite{singh:prb08} have also been suggested but
are unproven.

\subsection{Weak phonon-magnon coupling:}

In our Raman spectra (Figs.~\ref{raman:ht}c and d) we observed
small (1 or 2 \cm) increases or decreases in phonon frequencies
very near \tn. Note that this is observed for several different
phonon symmetries. The symmetry-independence of the phonon-magnon
coupling implies an interaction of form $P$$^{2}$$M$$^{2}$ (where
$P$ and $M$ are the polarization and magnetization, respectively)
in the free energy, as first suggested for magnetoelectrics by
Smolenskii and Chupis \cite{smolenkii:SPU82}. In general the
coupling of phonons and magnons can occur through several
different microscopic physical models: The Torrance-Slonczewski
model \cite{torrance:prb70} involves modulation of the crystal
field at the spin site by the optical phonon and is significant
for ions with unquenched orbital angular momenta, such as Co$^{\rm
+2}$ or Fe$^{\rm +2}$; the model of Buyers {\it et al}.
\cite{buyers:pr} is an angular momentum coupling of spins in
octahedra where the optical phonon eigenvector is rotation-like as
viewed from the magnetic ion. However, such models do not give
frequency anomalies near \tn\ like those we observed in \bfo.  A
rather detailed model of magneto-capacitance was given by Fox {\it
et al}. \cite{fox:prb} for BaMnF$_{4}$ near \tn, and related
models by Scott \cite{scott:pr77} and by Glass {\it et al}.
\cite{glass:ssp77} for the BaMF$_{4}$ family near T(2D), the
two-dimensional spin ordering temperature [typically ca. 3\tn\ in
that family]. The free energy of Fox {\it et al}. can be defined
as:
\begin{equation}
G = f(L^{2}, L_{z}^{2}) + B M^{2} + (b_{0} +b_{1}p +
b_{2}p^{2})M_{x}L_{z} + ..., \label{equn1}
\end{equation}
where  P is along $y$, the polar axis and $z$ is the sublattice
magnetization direction; $L$ is (1/2)g$\mu$ S$^{1/2}$
(S+1)$^{1/2}$ [ $\sum$ (Sj up - Sj down)] and M is the weak
magnetization M = g$\mu$ S$^{1/2}$ (S+1)$^{1/2}$ [$\sum$
(Sj~up~+~Sj~down)]. Note that $p$ is not the total polarization P
but only the part induced by the magnetoelectric coupling:
$P$~=~\pr~+~$p$ and it was carried out to second order in
polarization $P$, sublattice magnetization $M$, and weak
ferromagnetization $L$, with the result that the
magneto-capacitance varies with temperature as
(b$_{0}$~-~b$_{1}$b$_{2}$)L$_{2}$(T), where b$_{0}$, b$_{1}$, and
b$_{2}$ are respectively the coefficients of magnetoelectric free
energy terms independent of, linear in, and quadratic in
polarization P. The authors noted that the mean-field theory,
although generally not satisfactory for magnetic transitions,
works well for weakly canted ferromagnets because the expansion
parameter $L$ is small at all temperatures. Note that the sign of
the magneto-capacitance term can be positive or negative depending
upon the magnitude of (b$_{0}$~-~b$_{1}$b$_{2}$).  Because they
used mean field theory, their work neglected the small term near
\tn\ due to fluctuations considered below.

Although the second-order theory of Fox {\it et al}.
\cite{fox:prb} was satisfactory for describing all the data in
BaMnF$_{4}$ near \tn, it is not sufficient for phonon behavior in
\bfo.  In this case it is necessary to go to fourth order in L.
The reasons are explained by Nugroho {\it et al}.
\cite{nugroho:prb07} in their work on YbMnO$_{3}$. In this case
the key term in the free energy is of the form gP$^{2}$L$^{2}$,
which for weak coupling gives an explicit interaction of electric
field to $L$ that results in a magnetocapacitance of
\begin{equation}
(g^{2}P^{2}/kT) \int [<L^{2}(x)L^{2}(0)> - <L^{2}>^{2}] dx.
 \label{equn2}
\end{equation}
Although this fourth-order term is higher order than the terms in
g$<$L$^{2}$$>$ considered by Fox {\it et al}. \cite{fox:prb}, it
is singular at \tn, because it is proportional to the cube of the
correlation length ($\eta$) that diverges at \tn.

The result is that the phonons in Raman effect in \bfo\ of any
symmetry will be expected to have small anomalies in their
frequencies at \tn. These small dips or jumps will be proportional
to temperature $t$$^{(\alpha-1)}$ \cite{nugroho:prb07}
\cite{mostovoy:2007}, where $t$ is reduced temperature,
$t$~=~[\tn-$T$]/\tn and $\alpha$ is the critical exponent
describing divergence of the specific heat \cite{munoz:prb2000}.
Since the $\alpha$ is typically small, the phonon frequencies
should vary approximately as \tn/[\tn-T] near \tn\ and in
principle could be used to evaluate critical exponent $\alpha$.
However in the present work the data are too imprecise for this
chore, and even higher resolution would by insufficient due to
phonon damping. The bump in phonon frequencies and linewidth are
qualitatively predicted from the non-mean field theory of Nugroho
{\it et al}. \cite{nugroho:prb07}. However, their model does not
predict magnitudes for the height (increase in frequency) or width
(how near the transition the increase occurs) of the bump. No
anomaly at all is predicted by the mean-field theory of Fox {\it
et al}. \cite{fox:prb}, which does not consider terms in the free
energy introduced by Nugroho {\it et al}. \cite{nugroho:prb07}. A
similar behavior has been observed at 150~K (Figs.~\ref{raman:lt}b
and c) could be due to the change in magnetic ordering
\cite{pradhan:apl05}. As matter of coincidence the bottom
electrode SRO has a ferromagnetic phase transition at 150~K. Note
that none of these peaks is related to SRO and a modulated effect
is highly unlikely, but not impossible.

In summary, our Raman frequencies near \tn\ and 150~K show small
peaks or dips for all phonon modes that are qualitatively similar
to those predicted by Nugroho {\it et al}. \cite{nugroho:prb07},
implying a general interaction of form P$^{2}$L$^{2}$, and the
need for a fluctuation term neglected in the mean-field, weak
ferromagnetism model of Fox {\it et al} \cite{fox:prb}.

\section{CONCLUSION}
In conclusion, high quality epitaxial (001)BFO films have been
grown on (100) STO substrates using PLD. The XRD studies showed
that films are $c$-axis oriented with high degree of
crystallinity. The RT polarized Raman scattering of (001)BFO films
showed pseudo-orthorhombic monoclinic crystal structure contrary
to the rhombohedral and tetragonal symmetries reported earlier. We
observed the ferroelectric phase transition at around 820\dc\, and
no softening of Raman modes was observed at low frequencies, as in
BFO single crystals. The AFM-PM phase transition at around 370\dc\
caused some small changes in the phonon frequencies, linewidth,
and/or intensities of several low frequency modes, indicating ME
coupling in the material. A non-cubic $\gamma$-\bfo\ phase was
observed between 931-950\dc\ in the BFO thin films, in agreement
with the accepted \bfo\ phase diagram. The spectra also revealed
an evidence of a cubic $\delta$-phase around 1000\dc\ in thin
films that is not stable in powder and bulk.

\section*{ACKNOWLEDGEMENT}
The authors are grateful to Prof. Hans Schmid, University of
Geneva for supplying BFO single crystal for the comparative study
with the thin film and G. Catalan, University of Cambridge for his
useful comments and suggestions. We thank Brahim Dkhil for sharing
his Raman data on phonon anomalies in BFO single crystals near
140K prior to their publication.

The authors would like to acknowledge financial support from DOE
DE-FG02-08ER46526 grant. RP thanks Institute of Functional
Nanomaterial (IFN), UPR for financial support.
\section*{}

\end{document}